\documentclass{article}

\usepackage{PRIMEarxiv}

\usepackage[utf8]{inputenc} 
\usepackage[T1]{fontenc}    
\usepackage{hyperref}       
\usepackage{url}            
\usepackage{booktabs}       
\usepackage{amsfonts}       
\usepackage{nicefrac}       
\usepackage{microtype}      
\usepackage{lipsum}
\usepackage{fancyhdr}       
\usepackage{graphicx}       
\usepackage{lscape}
\usepackage{array}
\usepackage{algorithm}
\usepackage{algpseudocode}
\usepackage{caption}
\usepackage{amsmath}
\graphicspath{{media/}}     

\newcommand{\CenterRow}[2]{
  \dimen0=\ht\strutbox%
  \advance\dimen0\dp\strutbox%
  \multiply\dimen0 by#1%
  \divide\dimen0 by2%
  \advance\dimen0 by-.5\normalbaselineskip%
  \raisebox{-\dimen0}[0pt][0pt]{#2}}

\title{Scaling Laws for Moral Machine Judgment in Large Language Models
}

\author{
  Kazuhiro Takemoto \\
  Kyushu Institute of Technology \\
  Iizuka, Fukuoka, Japan \\
  \texttt{takemoto.kazuhiro035@m.kyutech.ac.jp} \\
\AND
}

\begin{document}
\captionsetup[table]{skip=7pt}

\maketitle

\begin{abstract}
Autonomous systems increasingly require moral judgment capabilities, yet whether these capabilities scale predictably with model size remains unexplored. We systematically evaluate 75 large language model configurations (0.27B--1000B parameters) using the Moral Machine framework, measuring alignment with human preferences in life-death dilemmas. We observe a consistent power-law relationship with distance from human preferences ($D$) decreasing as $D \propto S^{-0.10\pm0.01}$ ($R^2=0.50$, $p<0.001$) where $S$ is model size. Mixed-effects models confirm this relationship persists after controlling for model family and reasoning capabilities. Extended reasoning models show significantly better alignment, with this effect being more pronounced in smaller models (size$\times$reasoning interaction: $p = 0.024$). The relationship holds across diverse architectures, while variance decreases at larger scales, indicating systematic emergence of more reliable moral judgment with computational scale. These findings extend scaling law research to value-based judgments and provide empirical foundations for artificial intelligence governance.
\end{abstract}


\section{Introduction}
Autonomous systems increasingly face ethical dilemmas requiring value-based 
judgments~\cite{anderson2011machine}. Self-driving vehicles must decide how to 
allocate risk in unavoidable collision scenarios~\cite{bonnefon2016social}, 
while medical artificial intelligence (AI) systems navigate complex tradeoffs between patient autonomy 
and beneficence~\cite{ratti2025ethical}. The rapid advancement of large language 
models (LLMs) has accelerated their integration into such safety-critical 
decision-making roles~\cite{haltaufderheide2024ethics,yang2023llm4drive,nouri2024engineering,takemoto2024moral,takemoto2025moral}, 
making their moral judgment capabilities a central concern for AI safety and 
governance.

Recent work has established scaling laws across diverse cognitive capabilities. 
Model performance improves with increased parameters in language 
modeling and understanding~\cite{kaplan2020scaling,hoffmann2022training}. Many sophisticated capabilities 
emerge spontaneously through scale alone, without explicit training~\cite{wei2022emergent}. 
These observations raise the question of whether moral judgment capabilities 
similarly scale with model size, following patterns analogous to other cognitive 
domains.

The Moral Machine experiment~\cite{awad2018moral} provides a framework for 
investigating this question. This study collected 40 million decisions from 
participants across 233 countries on autonomous vehicle dilemmas, measuring how 
moral factors including age, social status, species, and legality influence 
choices. Rather than aggregating preferences into single correct answers, the 
framework quantifies the relative strength of different moral considerations, 
yielding preference distributions that capture value pluralism. This design 
enables measurement of alignment with human preference patterns through post-hoc 
statistical analysis, assessing moral intuitions less likely to have been 
explicitly targeted during model training or alignment procedures. Our previous 
work~\cite{takemoto2025moral} used this framework to provide 
initial evidence for scaling in moral judgment, observing improved alignment 
with human preferences in larger models. However, that analysis was limited to 
a small number of available models. The rapid proliferation of LLMs spanning 
orders of magnitude in size now enables systematic investigation of whether 
moral alignment follows predictable scaling laws.

Several recent benchmarks evaluate ethical reasoning in LLMs using crowdsourced 
labels to establish correct answers. ETHICS~\cite{hendrycks2021aligning} assesses 
knowledge across multiple ethical frameworks including utilitarianism and 
deontology. Moral Stories~\cite{emelin2021moral} evaluates moral reasoning through 
narrative understanding. Social Chemistry~\cite{forbes2020social} measures 
understanding of social norms and rules of thumb. These benchmarks provide 
valuable tools for assessing moral knowledge and reasoning about established 
norms. However, for real-world dilemmas faced by autonomous systems—such as 
unavoidable collision scenarios or medical tradeoffs—where no objective ground 
truth exists and multiple reasonable perspectives coexist, frameworks that 
preserve preference distributions offer complementary advantages. The Moral 
Machine's focus on preference alignment rather than classification accuracy 
makes it well-suited for studying how moral judgment capabilities scale with 
model size in contexts where human values are pluralistic.

We systematically investigate scaling relationships in moral judgment across 75 
LLM configurations spanning 0.27B to 1000B parameters, representing a substantial 
expansion from prior work. Our analysis reveals a consistent power-law relationship where larger models 
systematically demonstrate improved alignment with human moral preferences. Mixed-effects 
models confirm this relationship persists after controlling for model family and 
reasoning capabilities. These findings demonstrate that moral judgment follows 
predictable scaling patterns, extending scaling law research to value-based 
judgments and providing quantitative foundations for AI governance.

\section{Methods}
\subsection{Large Language Models}
We evaluated 75 model configurations spanning 0.27B to 1000B parameters using 
the evaluation framework established in our previous work~\cite{takemoto2024moral,
takemoto2025moral}. Complete methodological details including prompt formats, 
response coding procedures, and validation against the original Moral Machine 
human data are provided in those papers. The 75 configurations include different 
model sizes within families (e.g., Llama 3 8B, 70B), versions (e.g., DeepSeek 
V3, V3.1, V3.2), and reasoning modes (e.g., standard vs. extended reasoning models including reasoning-focused and thinking-mode models).

For each model family, we prioritized instruction-tuned or chat-optimized variants 
over base models, as base models often produce unreliable responses to conversational 
prompts. This ensures that observed scaling relationships reflect genuine moral 
judgment capabilities rather than artifacts of poor instruction-following.
The model set encompasses diverse architectures and training approaches, from 
both open-weight sources (Llama, Gemma, Qwen, Mistral, DeepSeek) and proprietary 
APIs (GPT and Claude).
For open-weight models, we used officially reported parameter counts. For 
proprietary models, we used parameter estimates from published literature~
\cite{abacha2024medec}, as official counts are not publicly disclosed.
Complete model specifications, release dates, and evaluation results 
are provided in Table~\ref{tab:model_metadata_distance}.

All models received identical prompts with default sampling parameters.
Responses were coded following the procedures detailed in~\cite{takemoto2024moral}.

\subsection{Moral Machine Framework}
We employed the Moral Machine framework~\cite{awad2018moral}, a validated 
instrument for measuring moral preferences in autonomous vehicle dilemmas. 
Following the methodology in~\cite{takemoto2024moral}, we generated 10,000 
distinct scenarios by systematically varying nine moral factors: age (young vs. 
elderly), gender (male vs. female), social status (high vs. low), physical 
fitness (fit vs. large), species (human vs. pet), legality (legal vs. illegal 
crossing), number of characters (more vs. fewer), intervention type (swerve vs. 
stay), and whether characters are passengers or pedestrians.

Standard models and thinking-mode models with thinking disabled evaluated all 
10,000 scenarios. Reasoning-focused models and thinking-mode models with thinking 
enabled evaluated the first 5,000 scenarios due to longer inference times. 
Complete prompt templates, response coding procedures, and scenario generation 
details are provided in~\cite{takemoto2024moral}.

\subsection{Evaluation Metrics}
Following~\cite{takemoto2024moral}, we quantified alignment between model 
judgments and human preferences using the Average Marginal Component Effect 
(AMCE)~\cite{hainmueller2014causal}. AMCE measures the causal influence of 
each moral factor on decision-making by estimating the average change in choice 
probability when a factor level changes (e.g., from young to elderly) while 
holding other factors constant. For each model, we computed AMCE values for 
all nine moral factors, yielding a 9-dimensional preference vector (Table~\ref{tab:amce_values}). 

We quantified overall alignment as the Euclidean distance between each model's 
AMCE vector and the aggregate human AMCE vector derived from the original Moral 
Machine experiment~\cite{awad2018moral}. Lower distances indicate greater 
correspondence to human moral preferences. Complete details on AMCE computation 
and human preference vectors are provided in~\cite{takemoto2024moral}.

\subsection{Statistical Analysis}
We analyzed the relationship between model size (parameter count) and alignment 
metrics using Spearman rank correlation to account for potential non-linear 
relationships and outliers. To test for power-law scaling, we fit log-linear 
models ($\log_{10} D \sim \log_{10} S$) and compared goodness of fit ($R^2$) 
against alternative functional forms including linear ($D \sim S$), logarithmic 
($D \sim \log_{10} S$), and exponential models ($\log_{10} D \sim S$). The 
power-law exponent $\alpha$ and its standard error were estimated from the slope 
of the log-linear regression.

To assess whether the observed scaling relationship reflects effects of model 
size rather than correlated confounding factors, we employed linear mixed-effects 
models with model family (DeepSeek, Llama, Gemma, Qwen, Other) as random effects, 
including family-specific intercepts and slopes for model size. We included two additional predictors: (1) release date, converted to 
a continuous variable (years) to capture temporal trends and control 
for potential scale-independent improvements in model capabilities, 
such as advances in training methodology, architecture design, or 
alignment procedures; (2) reasoning capability, coded 
as a binary variable indicating whether models employ extended reasoning processes 
(reasoning-focused models and thinking-mode variants = 1, standard models = 0). 
We compared nested models using likelihood ratio tests and information criteria 
(AIC, BIC): (A) model size only, (B) model size and release date, (C) model 
size, release date, and reasoning capability, and (D) model size, release 
date, reasoning capability, and size$\times$reasoning interaction.

All statistical analyses were performed using R software (version 4.4.2) with 
the lme4~\cite{bates2015lme4} and lmerTest~\cite{kuznetsova2017lmertest} packages. 
Mixed-effects models were fitted using restricted maximum likelihood (REML) for 
parameter estimation and maximum likelihood (ML) for model comparison.

\section{Results}
\subsection{Overall Scaling Relationship}
\begin{figure}[t]
\centering
\includegraphics[width=0.9\textwidth]{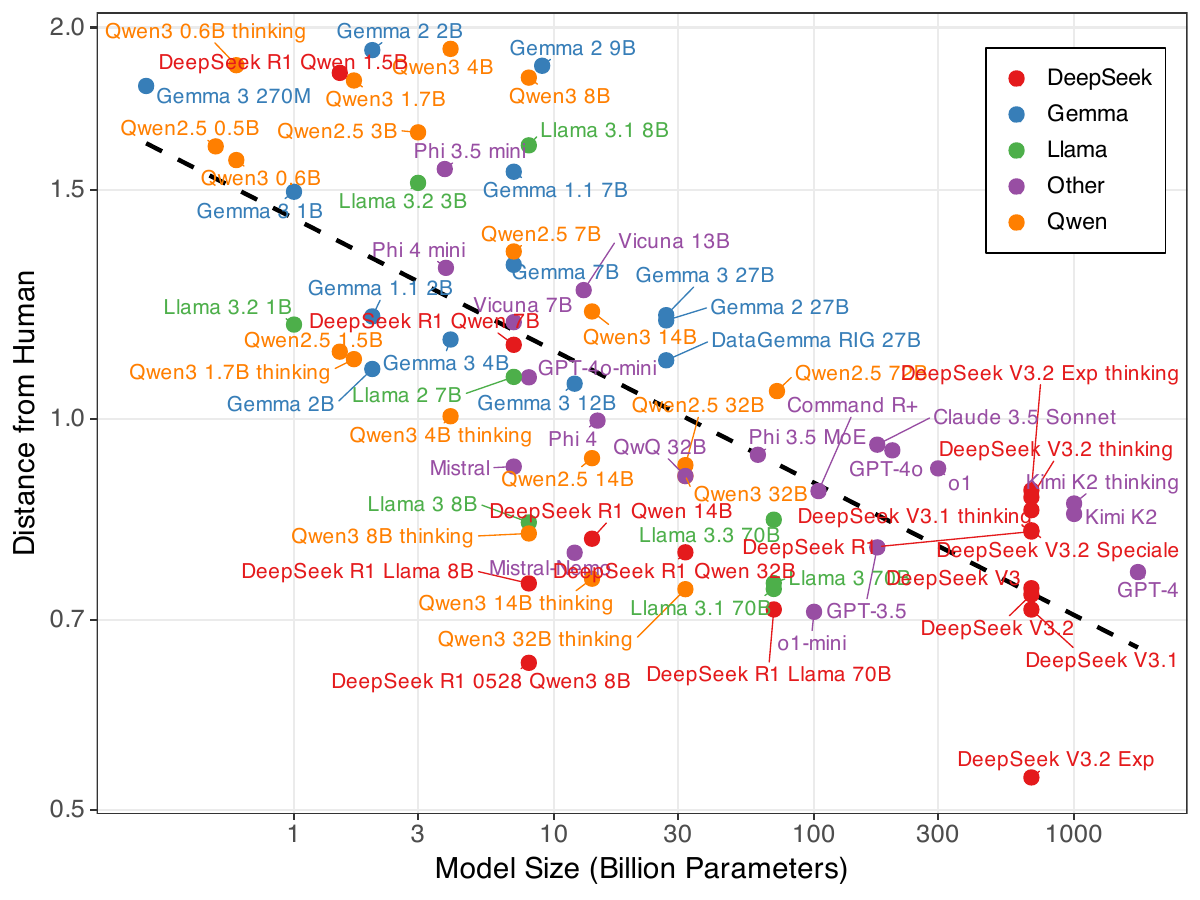} 
\caption{Scaling relationship between model size ($S$) and moral judgment 
alignment with human preferences (distance from human, $D$). Each point represents one LLM, 
colored by model family. The dashed line shows the fitted power-law 
relationship ($D \propto S^{-0.10\pm0.01}$)}
\label{fig:figure_logDlogS}
\end{figure}

We first examined the relationship between model size and moral judgment alignment 
across all 75 model configurations (Figure~\ref{fig:figure_logDlogS}). We observed a 
strong negative correlation (Spearman $\rho = -0.73$, $p = 1.8 \times 10^{-13}$), 
indicating that larger models systematically align better with human moral 
preferences. The relationship follows a power-law pattern $D \propto S^{-\alpha}$, 
where $D$ represents Euclidean distance from human preferences, $S$ denotes model 
size in parameters, and $\alpha = 0.10 \pm 0.01$. This power-law model achieved 
superior fit ($R^2 = 0.50$) compared to alternative functional forms: linear 
($R^2 = 0.16$), logarithmic ($R^2 = 0.47$), and exponential models ($R^2 = 0.19$). 
The power-law formulation outperforms the next-best model (logarithmic) by 
$\Delta R^2 = 0.03$, and substantially exceeds linear and exponential forms. 
The consistency of this relationship across 75 diverse models spanning three 
orders of magnitude in size (0.27B--1000B parameters) suggests a scaling law for moral judgment capabilities.

Beyond improved mean alignment, visual inspection of Figure~\ref{fig:figure_logDlogS} 
reveals that variance in alignment decreases with model size. Smaller models 
exhibit substantial variability in performance, while larger models cluster more 
tightly around the power-law trend, indicating more reliable and consistent moral 
judgment capabilities at scale.

\subsection{Robustness to Confounding Factors}
To assess whether the observed scaling relationship reflects genuine 
effects of model size rather than confounding factors, we fitted linear 
mixed-effects models with model family (DeepSeek, Llama, Gemma, Qwen, 
Other) as random effects. This approach accounts for family-specific 
differences in architecture and training approaches while estimating 
the effect of size. Model size correlates with several factors that 
could contribute to improved alignment, including computational budget, 
architectural sophistication, and advances in training methodology. 
While the first two can be partially controlled through family-specific 
random effects (as models within families share similar architectures 
and resource access), temporal improvements in model development are 
not directly observable. We used release date, converted to a continuous 
variable, as a proxy for potential scale-independent improvements over 
time, which could include advances in training practices, data curation, 
or alignment techniques. We compared nested models using likelihood 
ratio tests: (A) model size only, (B) model size and release date, 
(C) model size, release date, and reasoning capability, and (D) model size, 
release date, reasoning capability, and size$\times$reasoning interaction (Table~\ref{tab:model_comparison}).

\begin{table}[htbp]
\centering
\caption{Linear mixed-effects model comparison for moral alignment. All models 
include random intercepts and slopes for model family. Model comparison uses maximum likelihood estimation.}
\label{tab:model_comparison}
\begin{tabular}{llcccc}
\hline
Model & Fixed Effects & AIC & BIC & $\chi^2$ & $p$-value \\
\hline
A & Size & $-126.6$ & $-112.7$ & -- & -- \\
B & Size + Release date & $-124.7$ & $-108.5$ & 0.10 & 0.76 \\
C & Size + Release date + Reasoning & $-130.0$ & $-111.4$ & 7.24 & 0.007 \\
D & Size + Release date + Reasoning + Size$\times$Reasoning & $-133.0$ & $-112.1$ & 5.05 & 0.025 \\
\hline
\multicolumn{6}{l}{\small $\chi^2$ and $p$-values from likelihood ratio tests against previous model}
\end{tabular}
\end{table}

Model comparison revealed that adding release date did not improve fit 
(Model B vs. A: $\chi^2 = 0.10$, $p = 0.76$), suggesting that temporal 
factors in model development, which could include advances in training 
methodology, architectural innovations, or alignment procedures, do not 
substantially contribute to moral alignment beyond the effects captured 
by model size and family. However, adding reasoning capability 
significantly improved the model (Model C vs. B: $\chi^2 = 7.24$, 
$p = 0.007$). The full model including the size$\times$reasoning 
interaction provided further significant improvement (Model D vs. C: 
$\chi^2 = 5.05$, $p = 0.025$).

The final model (Model D; Table~\ref{tab:modelD_detail}) confirmed the power-law 
scaling relationship while controlling for family differences and temporal trends. 
Model size showed a significant negative effect ($\beta = -0.12$, SE $= 0.016$, 
$p < 0.001$), corresponding to a power-law exponent $\alpha = 0.12$ consistent 
with the overall analysis. Extended reasoning models showed significantly better 
alignment ($\beta = -0.16$, SE $= 0.044$, $p = 0.001$). Family-specific patterns 
(Figure~\ref{fig:family_scaling}) show predominantly negative relationships 
between size and alignment, though statistical power for detecting these 
relationships within individual families varies due to differences in sample 
size (8--19 models per family) and parameter range coverage. Random effects 
captured modest family-level variation (SD$_{\text{intercept}} = 0.019$, 
SD$_{\text{slope}} = 0.017$; Table~\ref{tab:modelD_detail}), with the 
mixed-effects approach providing robust estimates by pooling information across 
families while accounting for family-specific heterogeneity. Critically, the 
size$\times$reasoning interaction was positive and significant ($\beta = 0.057$, 
SE $= 0.025$, $p = 0.024$), indicating that the advantage of extended reasoning 
capabilities is more pronounced in smaller models (Figure~\ref{fig:reasoning_effect}). 
This suggests that architectural innovations in reasoning provide complementary 
pathways to improved alignment, with particularly strong benefits when 
computational scale is limited, while very large models may already capture 
much of this reasoning capability through scale alone.
Comparison of coefficient estimates across all nested models is provided in 
Table~\ref{tab:all_models}.

Analysis of residuals after controlling for size showed no correlation 
with release date (Spearman $\rho = 0.047$, $p = 0.69$; 
Figure~\ref{fig:temporal}; Table~\ref{tab:all_models}) 
further supporting that scale-independent temporal improvements in 
model development do not substantially contribute to alignment beyond 
scale and reasoning capabilities.
This null result should be interpreted 
cautiously given the limited temporal range of our dataset (95\% of models 
released in 2024--2025), which may constrain our ability to detect gradual 
improvements in training data quality over time.

\section{Discussion}
Our findings establish that computational scale systematically predicts moral 
judgment capabilities in LLMs. Across 75 model configurations spanning three 
orders of magnitude (0.27B--1000B parameters), we observed a consistent power-law 
relationship ($D \propto S^{-0.10 \pm 0.01}$, $R^2=0.50$, $p<0.001$), 
where distance from human moral preferences decreases predictably with model size. 
This relationship proved robust to potential confounding factors: mixed-effects 
models controlling for model family confirmed that the scaling pattern persists 
across diverse architectures and training approaches. Family-specific analyses 
revealed predominantly negative relationships between size and alignment across 
all major model families (DeepSeek, Llama, Gemma, Qwen), demonstrating that 
the power-law pattern is not driven by any single architectural approach or 
training methodology.

While computational scale represents an important factor in moral alignment, 
additional factors contribute independently. Extended reasoning models demonstrated 
significantly better alignment ($\beta = -0.16$, $p = 0.001$). Critically, this 
effect varies with model size: the size$\times$reasoning interaction ($\beta = 0.057$, 
$p = 0.024$) indicates that extended reasoning provides greater relative benefits 
in smaller models, while very large models already capture much of this capability 
through scale alone. This indicates that architectural or training innovations 
enabling extended reasoning, such as chain-of-thought processing~\cite{wei2022chain,wang2023selfconsistency} 
and iterative refinement~\cite{madaan2023self}, enhance moral judgment, with 
this enhancement being particularly pronounced when computational scale is limited. 
In contrast, temporal factors in model development, proxied by release 
date, did not contribute substantially to alignment after controlling 
for size and reasoning capability ($p = 0.76$).
This null result suggests that within our dataset, the variance in alignment is better explained by computational scale than by temporal trends in model development, possibly because improvements in moral alignment over time are largely mediated by increases in model size rather than scale-independent advances in training or alignment procedures. This interpretation should be treated cautiously, however, given the limited temporal range of our dataset (95\% of models released in 2024--2025) and the presence of other confounding factors discussed below.

The modest power-law exponent ($\alpha = 0.10$) reveals an important characteristic 
of moral judgment capabilities: they scale slowly with model size. A tenfold 
increase in parameters reduces distance from human preferences by only approximately 
21\%, indicating that moral judgment represents a particularly challenging emergent 
capability requiring substantial computational scale to approach human-level 
performance. This gradual scaling contrasts with steeper improvements observed 
in some other domains and, while subject to the interpretation of our distance-based 
metric, suggests fundamental computational constraints on acquiring human-like 
moral intuitions.

For practical deployment, this slow scaling implies that achieving substantial improvements in moral alignment requires order-of-magnitude increases in model size or complementary approaches such as extended reasoning architectures. This raises a fundamental question for AI safety: whether pure parameter scaling alone represents a sufficient or efficient path toward human-level moral alignment. Our findings suggest that it may not, and that architectural innovations such as extended reasoning are not merely complementary but may be strictly necessary for safety-critical deployments where computational resources are constrained. Extended reasoning provides particularly strong benefits in resource-constrained settings where deploying very large models is impractical, offering an alternative pathway to improved alignment without proportional increases in parameter count.

Computational scale also offers a distinct benefit beyond mean alignment: larger models exhibit reduced variance in performance, clustering more tightly around the power-law trend. This convergence could reflect two complementary mechanisms: larger models may be converging toward a more universal moral profile that reflects the aggregate diversity of their training data, or scale may primarily enhance internal consistency and predictability without necessarily implying convergence to a single shared value system. Distinguishing between these interpretations remains an important direction for future work. Regardless of the underlying mechanism, this increased reliability suggests that larger models offer more predictable moral reasoning, which is particularly valuable in safety-critical applications.

These findings extend scaling law research from tasks with objective ground truth 
to value-based judgments where correctness is defined by alignment with human 
preferences rather than factual accuracy. 
Previous scaling studies ~\cite{kaplan2020scaling,hoffmann2022training} have focused predominantly on capabilities with verifiable solutions and factual question answering, where performance can be evaluated against definitive correct answers.
Our demonstration that moral judgment follows analogous power-law 
patterns despite lacking objective ground truth suggests that scaling laws may 
represent a more general principle of capability emergence~\cite{wei2022emergent} in large language models. 
This extension has important theoretical implications: it indicates that models 
can acquire implicit knowledge of human value systems through exposure to training 
data reflecting diverse ethical perspectives, and that this acquisition process 
follows predictable quantitative relationships with computational scale.
The consistency of scaling patterns across model families suggests that computational scale represents an important factor in moral alignment. While architectural innovations such as extended reasoning provide complementary improvements, the relationship between model size and moral judgment quality follows a consistent power-law pattern across diverse implementation choices.

These findings provide quantitative foundations for risk-based deployment of LLMs 
in contexts requiring moral judgment. The power-law relationship enables prediction 
of alignment quality from model size, allowing system designers to make informed 
tradeoffs between computational costs and ethical reliability. For safety-critical 
applications where ethical failures carry substantial consequences, our results 
indicate that larger models or those enhanced with extended reasoning capabilities 
offer improved reliability. Extended reasoning represents a complementary pathway 
to enhanced alignment that does not require proportional increases in parameter 
count, with particularly strong benefits when computational scale is limited. 
Beyond mean alignment, the observed reduction in variance at larger scales implies 
more predictable performance, reducing the risk of outlier judgments that could 
undermine trust or violate ethical standards in high-stakes 
contexts~\cite{rudin2019stop,reinhardt2023trust,jiao2025navigating}.

Several limitations warrant consideration. Our evaluation uses Euclidean distance 
between model and human AMCE vectors as the primary alignment metric. While this 
metric provides mathematical rigor and enables quantitative comparison across 
models, the relationship between distance reductions and practical improvements 
in real-world moral judgment quality remains to be empirically validated. Future 
work should complement distance-based metrics with human evaluations to establish 
the practical significance of observed improvements.

Model size correlates with multiple factors including computational 
budget, architectural sophistication, and advances in training 
methodology. While our mixed-effects analyses control for model family 
and temporal trends in model development, fully isolating the causal 
effect of scale from all correlated factors requires controlled 
experiments that are impractical at the scale of contemporary LLMs. 
Training data for most models is not publicly disclosed, making it impossible to assess whether Moral Machine scenarios appeared in pretraining corpora. However, our evaluation scenarios were systematically generated by combining nine moral factors across multiple levels, producing a combinatorial space that is effectively infinite and thus unlikely to have been encountered verbatim in pretraining data. Simple memorization of specific scenario outcomes would therefore be insufficient to explain the observed scaling patterns. Furthermore, the consistency of scaling patterns across models from different organizations suggests systematic relationships rather than memorization artifacts.

Our evaluation focuses on life-death dilemmas from the Moral Machine framework, 
representing one aspect of moral reasoning. We selected this framework because 
its preference-based evaluation approach, rather than correct-answer classification, 
reduces the risk of data contamination from training corpora while enabling 
measurement of spontaneous moral intuitions. Whether similar scaling relationships 
exist for other ethical dimensions such as fairness, autonomy, and virtue ethics 
remains an important future direction.
The Moral Machine dataset predominantly reflects preferences from participants in Western countries, and the alignment metric used in this study refers specifically to correspondence with this aggregate preference distribution rather than to any universal moral standard. Scaling relationships observed here may therefore differ if models were evaluated against specific cultural or regional moral sub-clusters. Cross-cultural validation~\cite{vida2024decoding,jin2024language} is essential to assess the generalizability of our findings across diverse value systems.
Finally, different prompting strategies might yield 
quantitatively different relationships~\cite{oh2025robustness,kim2025exploring}, 
though we expect directional trends to remain consistent.

Despite the limitations discussed above, our results indicate that computational 
scale represents an important factor in moral alignment. The observed power-law 
relationship provides empirical evidence that model size systematically predicts 
alignment with human moral preferences, suggesting that human-aligned moral 
capabilities emerge with computational scale. These findings offer initial 
quantitative insights for AI governance \cite{dafoe2018ai,batool2025ai} and inform risk assessment in deploying AI systems \cite{novelli2024taking,novelli2024ai} requiring value-based judgments.

\section*{Acknowledgments}
This research was funded by the JSPS KAKENHI (grant number 21H03545).

\bibliographystyle{unsrt}  
\bibliography{references}  

@article{
wei2022emergent,
title={Emergent abilities of large language models},
author={Jason Wei and Yi Tay and Rishi Bommasani and others},
journal={Transactions on Machine Learning Research},
issn={2835-8856},
year={2022},
url={https://openreview.net/forum?id=yzkSU5zdwD},
note={Survey Certification}
}

@article{kaplan2020scaling,
  title={Scaling laws for neural language models},
  author={Kaplan, Jared and McCandlish, Sam and Henighan, Tom and others},
  journal={arXiv preprint arXiv:2001.08361},
  year={2020}
}

@inproceedings{hoffmann2022training,
author = {Hoffmann, Jordan and Borgeaud, Sebastian and Mensch, Arthur and others},
title = {Training compute-optimal large language models},
year = {2022},
booktitle = {Proceedings of the 36th International Conference on Neural Information Processing Systems (NeurIPS)},
articleno = {2176},
numpages = {15},
}

@book{anderson2011machine,
  title={Machine ethics},
  author={Anderson, Michael and Anderson, Susan Leigh},
  year={2011},
  publisher={Cambridge University Press}
}

@article{bonnefon2016social,
  title={The social dilemma of autonomous vehicles},
  author={Bonnefon, Jean-Fran{\c{c}}ois and Shariff, Azim and Rahwan, Iyad},
  journal={Science},
  volume={352},
  number={6293},
  pages={1573--1576},
  year={2016},
  publisher={American Association for the Advancement of Science}
}

@article{ratti2025ethical,
  title={Ethical and social considerations of applying artificial intelligence in healthcare—a two-pronged scoping review},
  author={Ratti, Emanuele and Morrison, Michael and Jakab, Ivett},
  journal={BMC Medical Ethics},
  volume={26},
  number={1},
  pages={68},
  year={2025},
  publisher={Springer}
}

@article{awad2018moral,
  title={The moral machine experiment},
  author={Awad, Edmond and Dsouza, Sohan and Kim, Richard and others},
  journal={Nature},
  volume={563},
  number={7729},
  pages={59--64},
  year={2018},
  publisher={Nature Publishing Group UK London}
}

@article{haltaufderheide2024ethics,
  title={The ethics of ChatGPT in medicine and healthcare: a systematic review on Large Language Models (LLMs)},
  author={Haltaufderheide, Joschka and Ranisch, Robert},
  journal={NPJ digital medicine},
  volume={7},
  number={1},
  pages={183},
  year={2024},
  publisher={Nature Publishing Group UK London}
}

@article{takemoto2024moral,
  title={The moral machine experiment on large language models},
  author={Takemoto, Kazuhiro},
  journal={Royal Society Open Science},
  volume={11},
  number={2},
  pages={231393},
  year={2024},
  publisher={The Royal Society}
}

@inproceedings{emelin2021moral,
  title={Moral stories: Situated reasoning about norms, intents, actions, and their consequences},
  author={Emelin, Denis and Le Bras, Ronan and Hwang, Jena D and Forbes, Maxwell and Choi, Yejin},
  booktitle={Proceedings of the 2021 Conference on Empirical Methods in Natural Language Processing},
  pages={698--718},
  year={2021}
}

@inproceedings{forbes2020social,
  title={Social Chemistry 101: Learning to Reason about Social and Moral Norms},
  author={Maxwell Forbes and Jena D. Hwang and Vered Shwartz and Maarten Sap and Yejin Choi},
  booktitle={Conference on Empirical Methods in Natural Language Processing},
  year={2020},
  url={https://api.semanticscholar.org/CorpusID:226226666}
}

@inproceedings{
hendrycks2021aligning,
title={Aligning {\{}AI{\}} With Shared Human Values},
author={Dan Hendrycks and Collin Burns and Steven Basart and Andrew Critch and Jerry Li and Dawn Song and Jacob Steinhardt},
booktitle={International Conference on Learning Representations},
year={2021},
url={https://openreview.net/forum?id=dNy_RKzJacY}
}

@article{takemoto2025moral,
  title={Large-scale moral machine experiment on large language models},
  author={Zaim bin Ahmad, Muhammad Shahrul and Takemoto, Kazuhiro},
  journal={PloS One},
  volume={20},
  number={5},
  pages={e0322776},
  year={2025},
  publisher={Public Library of Science San Francisco, CA USA}
}

@article{abacha2024medec,
  title={Medec: A benchmark for medical error detection and correction in clinical notes},
  author={Abacha, Asma Ben and Yim, Wen-wai and Fu, Yujuan and others},
  journal={arXiv preprint arXiv:2412.19260},
  year={2024}
}

@article{hainmueller2014causal,
  title={Causal inference in conjoint analysis: Understanding multidimensional choices via stated preference experiments},
  author={Hainmueller, Jens and Hopkins, Daniel J and Yamamoto, Teppei},
  journal={Political analysis},
  volume={22},
  number={1},
  pages={1--30},
  year={2014},
  publisher={Cambridge University Press}
}

@article{oh2025robustness,
  title={Robustness of large language models in moral judgements},
  author={Oh, Soyoung and Demberg, Vera},
  journal={Royal Society Open Science},
  volume={12},
  number={4},
  pages={241229},
  year={2025},
  publisher={The Royal Society}
}

@inproceedings{vida2024decoding,
  title={Decoding multilingual moral preferences: Unveiling LLM's biases through the Moral Machine experiment},
  author={Vida, Karina and Damken, Fabian and Lauscher, Anne},
  booktitle={Proceedings of the AAAI/ACM Conference on AI, Ethics, and Society},
  volume={7},
  pages={1490--1501},
  year={2024}
}

@article{batool2025ai,
  title={AI governance: a systematic literature review},
  author={Batool, Amna and Zowghi, Didar and Bano, Muneera},
  journal={AI and Ethics},
  pages={1--15},
  year={2025},
  publisher={Springer}
}

@Article{bates2015lme4,
    title = {Fitting Linear Mixed-Effects Models Using {lme4}},
    author = {Douglas Bates and Martin M{\"a}chler and Ben Bolker and
      Steve Walker},
    journal = {Journal of Statistical Software},
    year = {2015},
    volume = {67},
    number = {1},
    pages = {1--48},
    doi = {10.18637/jss.v067.i01},
  }

@Article{kuznetsova2017lmertest,
    title = {{lmerTest} Package: Tests in Linear Mixed Effects Models},
    author = {Alexandra Kuznetsova and Per B. Brockhoff and Rune H. B.
      Christensen},
    journal = {Journal of Statistical Software},
    year = {2017},
    volume = {82},
    number = {13},
    pages = {1--26},
    doi = {10.18637/jss.v082.i13},
  }

@article{novelli2024taking,
  title={Taking AI risks seriously: a new assessment model for the AI Act},
  author={Novelli, Claudio and Casolari, Federico and Rotolo, Antonino and Taddeo, Mariarosaria and Floridi, Luciano},
  journal={Ai \& Society},
  volume={39},
  number={5},
  pages={2493--2497},
  year={2024},
  publisher={Springer}
}

@article{novelli2024ai,
  title={AI risk assessment: a scenario-based, proportional methodology for the AI act},
  author={Novelli, Claudio and Casolari, Federico and Rotolo, Antonino and Taddeo, Mariarosaria and Floridi, Luciano},
  journal={Digital Society},
  volume={3},
  number={1},
  pages={13},
  year={2024},
  publisher={Springer}
}

@article{dafoe2018ai,
  title={AI governance: a research agenda},
  author={Dafoe, Allan},
  journal={Governance of AI Program, Future of Humanity Institute, University of Oxford: Oxford, UK},
  volume={1442},
  pages={1443},
  year={2018}
}

@article{reinhardt2023trust,
  title={Trust and trustworthiness in AI ethics},
  author={Reinhardt, Karoline},
  journal={AI and Ethics},
  volume={3},
  number={3},
  pages={735--744},
  year={2023},
  publisher={Springer}
}

@article{rudin2019stop,
  title={Stop explaining black box machine learning models for high stakes decisions and use interpretable models instead},
  author={Rudin, Cynthia},
  journal={Nature machine intelligence},
  volume={1},
  number={5},
  pages={206--215},
  year={2019},
  publisher={Nature Publishing Group UK London}
}

@article{wei2022chain,
  title={Chain-of-thought prompting elicits reasoning in large language models},
  author={Wei, Jason and Wang, Xuezhi and Schuurmans, Dale and Bosma, Maarten and Xia, Fei and Chi, Ed and Le, Quoc V and Zhou, Denny and others},
  journal={Advances in neural information processing systems},
  volume={35},
  pages={24824--24837},
  year={2022}
}

@article{jiao2025navigating,
  title={Navigating llm ethics: Advancements, challenges, and future directions},
  author={Jiao, Junfeng and Afroogh, Saleh and Xu, Yiming and Phillips, Connor},
  journal={AI and Ethics},
  pages={1--25},
  year={2025},
  publisher={Springer}
}

@inproceedings{
wang2023selfconsistency,
title={Self-Consistency Improves Chain of Thought Reasoning in Language Models},
author={Xuezhi Wang and Jason Wei and Dale Schuurmans and Quoc V Le and Ed H. Chi and Sharan Narang and Aakanksha Chowdhery and Denny Zhou},
booktitle={The Eleventh International Conference on Learning Representations },
year={2023},
url={https://openreview.net/forum?id=1PL1NIMMrw}
}

@article{madaan2023self,
  title={Self-refine: Iterative refinement with self-feedback},
  author={Madaan, Aman and Tandon, Niket and Gupta, Prakhar and Hallinan, Skyler and Gao, Luyu and Wiegreffe, Sarah and Alon, Uri and Dziri, Nouha and Prabhumoye, Shrimai and Yang, Yiming and others},
  journal={Advances in Neural Information Processing Systems},
  volume={36},
  pages={46534--46594},
  year={2023}
}

@inproceedings{nouri2024engineering,
  title={Engineering safety requirements for autonomous driving with large language models},
  author={Nouri, Ali and Cabrero-Daniel, Beatriz and T{\"o}rner, Fredrik and Sivencrona, H{\aa}kan and Berger, Christian},
  booktitle={2024 IEEE 32nd International Requirements Engineering Conference (RE)},
  pages={218--228},
  year={2024},
  organization={IEEE}
}

@article{jin2024language,
  title={Language model alignment in multilingual trolley problems},
  author={Jin, Zhijing and Kleiman-Weiner, Max and Piatti, Giorgio and others},
  journal={arXiv preprint arXiv:2407.02273},
  year={2024}
}

@article{yang2023llm4drive,
  title={LLM4Drive: A survey of large language models for autonomous driving},
  author={Yang, Zhenjie and Jia, Xiaosong and Li, Hongyang and Yan, Junchi},
  journal={arXiv preprint arXiv:2311.01043},
  year={2023}
}

@article{kim2025exploring,
  title={Exploring persona-dependent LLM alignment for the moral machine experiment},
  author={Kim, Jiseon and Kwon, Jea and Vecchietti, Luiz Felipe and Oh, Alice and Cha, Meeyoung},
  journal={arXiv preprint arXiv:2504.10886},
  year={2025}
}

\clearpage
\appendix
\setcounter{figure}{0}
\renewcommand{\thefigure}{S\arabic{figure}}

\setcounter{table}{0}
\renewcommand{\thetable}{S\arabic{table}}

\pagestyle{empty}

\section*{Supplementary Figures}
\begin{figure}[htbp]
\centering
\includegraphics[width=0.6\textwidth]{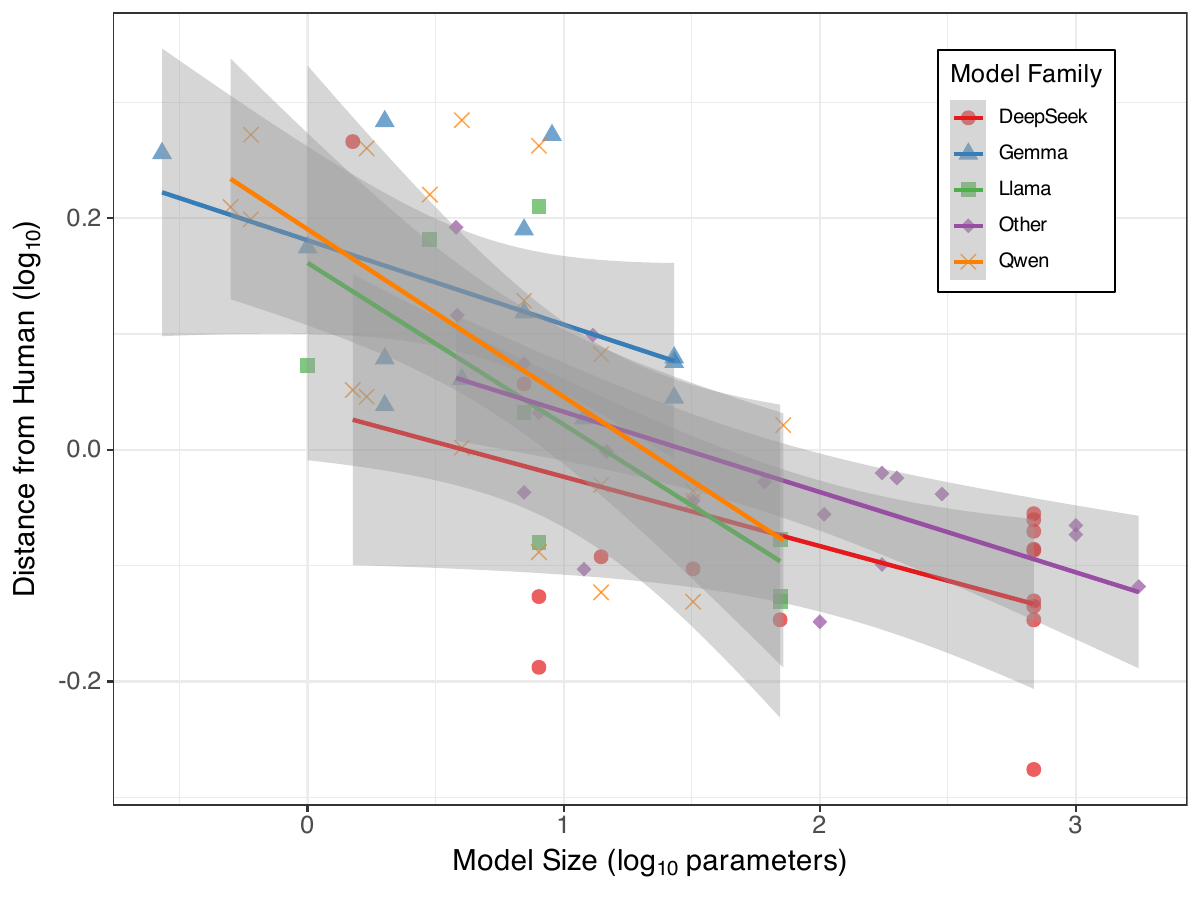}
\caption{Family-specific scaling relationships. Log-log plot showing 
the relationship between model size and distance from human preferences for 
each model family. Points represent individual models, lines show linear 
regression fits with 95\% confidence intervals. All families exhibit negative 
scaling relationships, demonstrating that the power-law pattern is not driven 
by any single architectural approach. Model families: DeepSeek (red circles), 
Gemma (blue triangles), Llama (green squares), Qwen (purple diamonds), Other 
(brown crosses).}
\label{fig:family_scaling}
\end{figure}

\begin{figure}[htbp]
\centering
\includegraphics[width=0.6\textwidth]{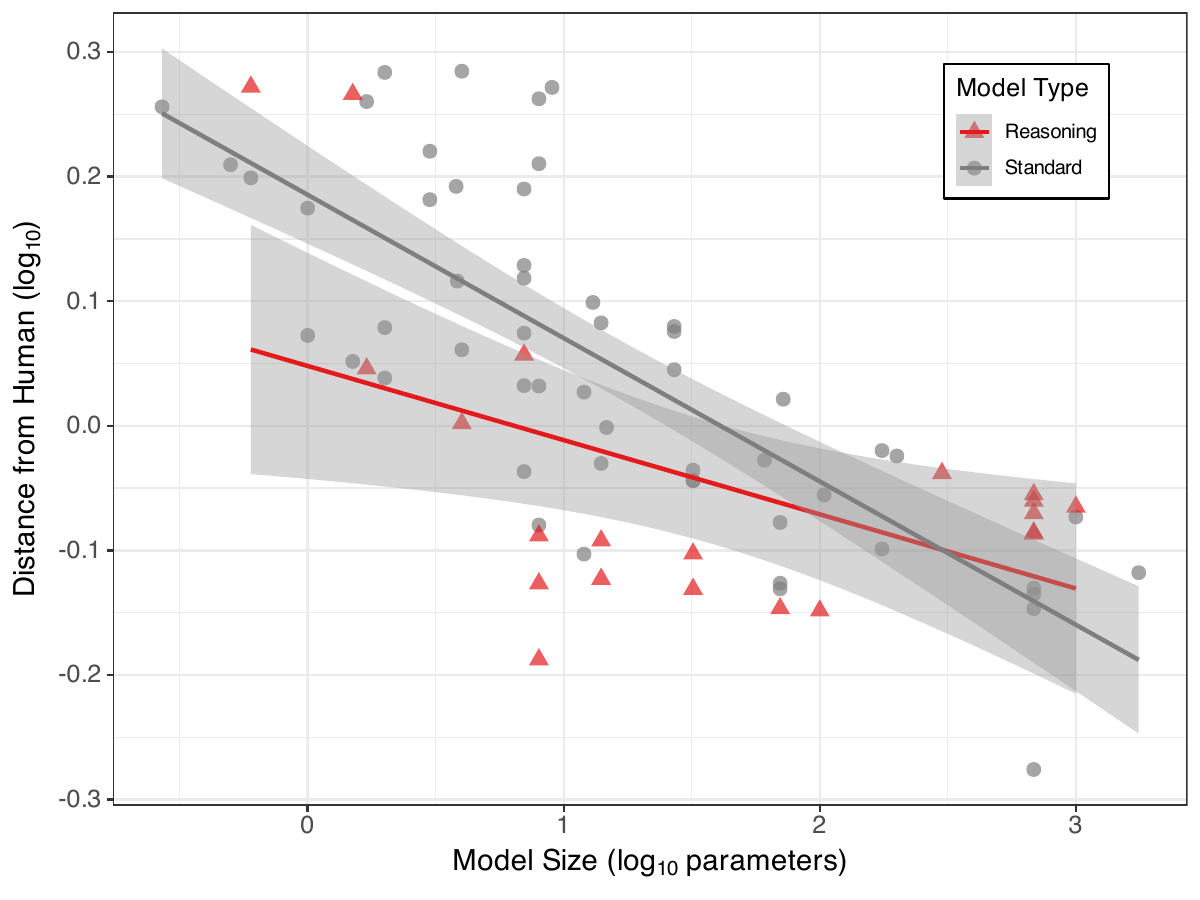}
\caption{Effect of extended reasoning capabilities on moral alignment.
Comparison of scaling relationships between standard models (gray circles) and 
extended reasoning models (red triangles). Extended reasoning models show 
systematically better alignment (lower distance) at comparable sizes, with 
the advantage being more pronounced in smaller models. The significant main 
effect of reasoning ($\beta = -0.16$, $p = 0.001$) and the size$\times$reasoning interaction ($\beta = 0.057$, $p = 0.024$) indicate that architectural 
innovations in reasoning provide particularly strong benefits when computational 
scale is limited. Lines show linear regression fits with 95\% confidence 
intervals.}
\label{fig:reasoning_effect}
\end{figure}

\begin{figure}[htbp]
\centering
\includegraphics[width=0.6\textwidth]{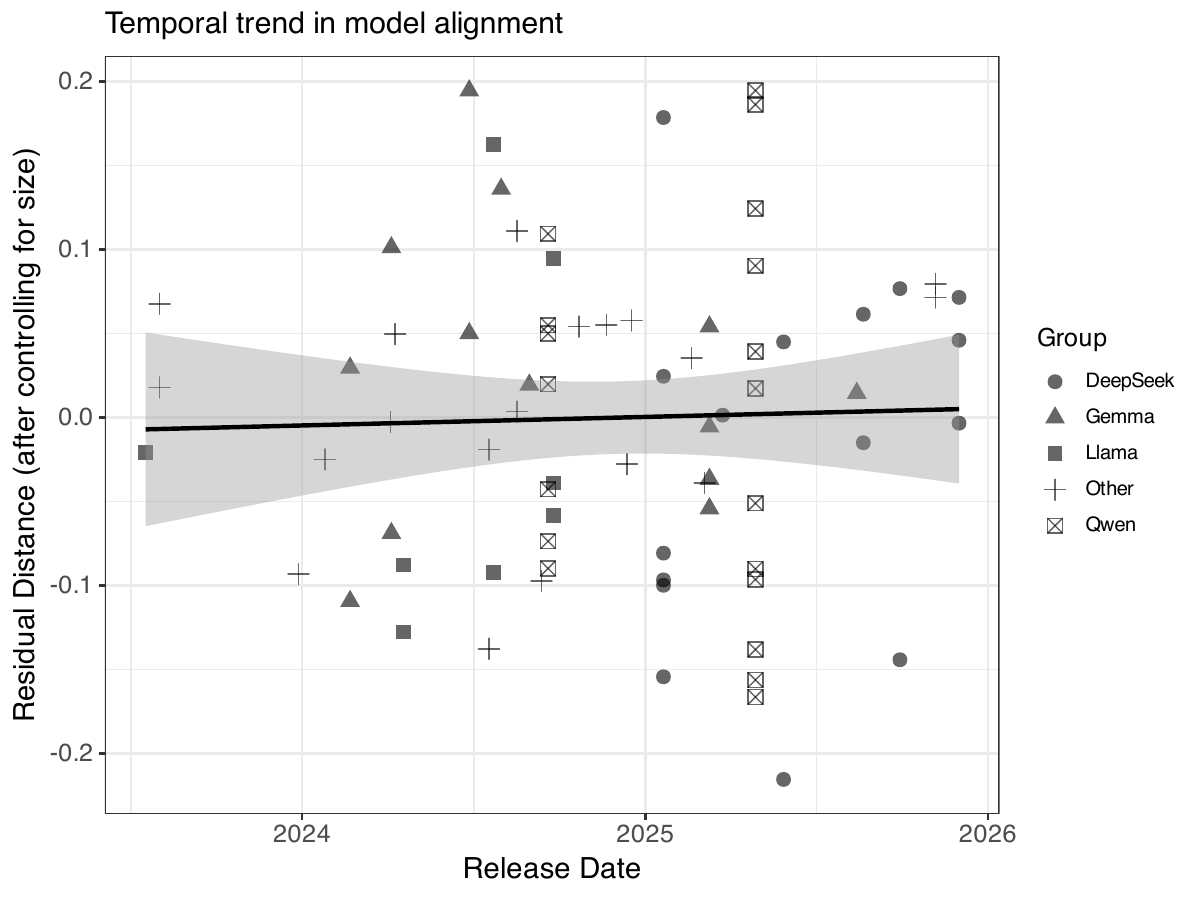}
\caption{
Temporal trends in moral alignment. Residual distance from human 
preferences (after controlling for model size) plotted against release 
date. The absence of a significant trend (Spearman $\rho = 0.047$, $p = 0.69$)
suggests that scale-independent temporal improvements in model 
development do not contribute substantially to alignment beyond the 
effects of scale and reasoning capabilities.
Points represent individual models, line shows linear 
regression fit with 95\% confidence interval.}
\label{fig:temporal}
\end{figure}

\clearpage

\section*{Supplementary Tables}

\begin{table}[htbp]
\centering
\caption{Model metadata and alignment results. This table provides 
key information for all 75 model configurations evaluated in this study. 
\textit{OpenWeight}: whether the model weights are publicly available (yes) or 
proprietary (no). \textit{Reasoning}: whether the model employs extended reasoning 
capabilities (yes = reasoning-focused models or thinking-mode variants; no = 
standard models). \textit{API ID}: official API identifier for proprietary models 
accessed via vendor APIs; blank for open-weight models evaluated locally. 
\textit{Size}: model size in billions of parameters; for proprietary models, 
values represent estimates from published literature~\cite{abacha2024medec} as 
official counts are not publicly disclosed. \textit{Release Date}: official 
release date in YYYY/MM/DD format. \textit{Distance}: Euclidean distance between 
the model's AMCE vector and the human AMCE vector (Table~\ref{tab:amce_values}), 
where lower values indicate closer alignment with human moral preferences. 
\textit{Valid Response Rate}: proportion of scenarios where the model provided a valid 
binary choice response that could be coded for analysis.
This is publicly available from our GitHub repository (\url{https://github.com/kztakemoto/mmllm/}).
}
\label{tab:model_metadata_distance}
\end{table}

\begin{table}[htbp]
\centering
\caption{Average Marginal Component Effect (AMCE) values for all models 
and human preferences. Each row represents one model configuration's (or human) 
9-dimensional preference vector, quantifying the causal influence of each moral 
factor on decision-making. AMCE values indicate the average change in choice 
probability when a factor level changes (e.g., from young to elderly) while 
holding other factors constant. Positive values indicate preference for the first 
level of each factor: young (vs. elderly), fit (vs. large), male (vs. female), 
swerve (vs. stay), legal crossing (vs. illegal), more characters (vs. fewer), 
passengers (vs. pedestrians), high status (vs. low), and human (vs. pet). Negative 
values indicate preference for the second level. Model size is reported in billions 
of parameters. The ``Human'' row represents aggregate preferences from the original 
Moral Machine experiment~\cite{awad2018moral} with 40 million decisions across 
233 countries.
This is publicly available from our GitHub repository (\url{https://github.com/kztakemoto/mmllm/}).
}
\label{tab:amce_values}
\end{table}


\begin{table}[htbp]
\centering
\caption{Mixed-effects model (Model D) parameter estimates. The model 
includes model size (log$_{10}$ parameters), release date, reasoning capability, 
and the size$\times$reasoning interaction as fixed effects, with random 
intercepts and slopes for model family. The model was fitted using restricted 
maximum likelihood (REML).}
\label{tab:modelD_detail}
\begin{tabular}{lrrrrr}
\hline
Parameter & Estimate & SE & df & $t$-value & $p$-value \\
\hline
\multicolumn{6}{l}{\textit{Fixed Effects}} \\
Intercept & 0.033 & 0.107 & 56.6 & 0.312 & 0.756 \\
Model Size (log$_{10}$) & $-0.120$ & 0.016 & 5.9 & $-7.397$ & $<$0.001 \\
Release Date & 0.033 & 0.023 & 58.4 & 1.458 & 0.150 \\
Reasoning & $-0.156$ & 0.044 & 40.3 & $-3.544$ & 0.001 \\
Size$\times$Reasoning & 0.057 & 0.025 & 56.3 & 2.313 & 0.024 \\
\hline
\multicolumn{6}{l}{\textit{Random Effects (Standard Deviations)}} \\
\multicolumn{6}{l}{Family: Intercept = 0.019} \\
\multicolumn{6}{l}{Family: Slope (Model Size) = 0.017} \\
\multicolumn{6}{l}{Correlation (Intercept, Slope) = $-1.00$} \\
\multicolumn{6}{l}{Residual = 0.091} \\
\hline
\multicolumn{6}{l}{\small Observations: 75; Groups (Family): 5} \\
\multicolumn{6}{l}{\small AIC = $-119.7$ (REML)} \\
\end{tabular}
\end{table}


\begin{table}[htbp]
\centering
\caption{Comparison of fixed effect estimates across nested mixed-effects 
models. All models include random intercepts and slopes for model family. Standard 
errors in parentheses. Models were fitted using REML for parameter estimation.}
\label{tab:all_models}
\begin{tabular}{lcccc}
\hline
Parameter & Model A & Model B & Model C & Model D \\
\hline
\multicolumn{5}{l}{\textit{Fixed Effects}} \\
Intercept & 0.142* & 0.093 & 0.006 & 0.033 \\
 & (0.027) & (0.110) & (0.106) & (0.107) \\
Model Size (log$_{10}$) & $-0.096$*** & $-0.096$*** & $-0.101$* & $-0.120$*** \\
 & (0.014) & (0.014) & (0.012) & (0.016) \\
Release Date & -- & 0.010 & 0.034 & 0.033 \\
 & & (0.023) & (0.023) & (0.023) \\
Reasoning & -- & -- & $-0.074$** & $-0.156$** \\
 & & & (0.027) & (0.044) \\
Size$\times$Reasoning & -- & -- & -- & 0.057* \\
 & & & & (0.025) \\
\hline
\multicolumn{5}{l}{\textit{Random Effects (SD)}} \\
Family: Intercept & 0.042 & 0.040 & 0.005 & 0.019 \\
Family: Slope (Size) & 0.014 & 0.009 & 0.005 & 0.017 \\
Residual & 0.095 & 0.095 & 0.094 & 0.091 \\
\hline
\multicolumn{5}{l}{\textit{Model Fit}} \\
AIC (REML) & $-125.1$ & $-119.5$ & $-120.2$ & $-119.7$ \\
\hline
\multicolumn{5}{l}{\small * $p < 0.05$; ** $p < 0.01$; *** $p < 0.001$}
\end{tabular}
\end{table}

\end{document}